\documentclass[a4paper,11pt]{article}
\usepackage{pos}
\usepackage{amsmath}
\usepackage{graphicx}
\usepackage{bm}

\newcommand{\be}{\begin{equation}}
\newcommand{\ee}{\end{equation}}
\newcommand{\bea}{\begin{eqnarray}}
\newcommand{\eea}{\end{eqnarray}}

\title{A novel computational paradigm for a precise determination of the 
       hadronic contribution to\\ $(g_\mu-2)$ from lattice QCD}

\ShortTitle{ML-HVP}

\author*[a,b]{Leonardo Giusti}
\author[a,b]{Mattia Dalla Brida}
\author[a,b]{Tim Harris}
\author[b]{Michele Pepe}

\affiliation[a]{Dipartimento di Fisica, Universit\`a di Milano-Bicocca\\ Piazza della Scienza 3, 
  I-20126 Milano, Italy}

\affiliation[b]{INFN, Sezione di Milano-Bicocca \\ Piazza della Scienza 3, 
I-20126 Milano, Italy}

\emailAdd{leonardo.giusti@mib.infn.it}
\emailAdd{mattia.dallabrida@mib.infn.it}
\emailAdd{tim.harris@mib.infn.it}
\emailAdd{michele.pepe@mib.infn.it}

\abstract{
The hadronic contribution to the muon anomalous magnetic moment $a_\mu=(g_\mu-2)/2$
has to be determined at the per-mille level for the Standard Model prediction
to match the expected final uncertainty of the ongoing E989 experiment.
That is 3 times better than the current precision from the dispersive approach, and 5-15
times smaller than the uncertainty based on the purely theoretical determinations from lattice QCD.
So far the stumbling-block is the large statistical error in the Monte Carlo evaluation of the
required correlation functions which can hardly be tamed by brute force. In this talk we
present our proposal to solve this problem by multi-level Monte Carlo integration, a technique
which reduces the variance of correlators exponentially in the distance of the fields.
We report the results of our feasibility tests for the computation of the Hadronic Vacuum Polarization
on a lattice with a linear extension of 3~fm, a spacing of 0.065 fm, and a pion mass of 270 MeV.
Indeed the two-level integration makes the contribution to the statistical
error from long-distances de-facto negligible by accelerating its inverse scaling with the cost
of the simulation. These findings establish multi-level Monte Carlo as a solid and efficient
method for a precise lattice determination of the hadronic contribution to $a_\mu$.
}

\FullConference{%
  40th International Conference on High Energy physics - ICHEP2020\\
  July 28 - August 6, 2020\\
  Prague, Czech Republic (virtual meeting)
}

\begin{document}
\maketitle

\section{Introduction}
\vspace{-0.25cm}

The measurement of the muon anomalous magnetic moment $a_\mu = 116 592 08.9(6.3) \times 10^{-10}$
by the E821 experiment has the remarkable precision of $0.54$ parts per million (ppm) \cite{Bennett:2006fi}, and
the on-going E989 experiment at FNAL is expected to reach the astonishing precision of $0.14$ ppm by the
end of its operation~\cite{Grange:2015fou}. The Standard Model (SM) prediction includes contributions
from five-loop Quantum Electrodynamics, two-loop Weak interactions, the Hadronic leading-order
Vacuum Polarization (HVP) and Hadronic Light-by-Light scattering (HLbL) \cite{Aoyama:2020ynm}.
The overall theoretical uncertainty is dominated by the hadronic part. So far, 
lacking precise purely theoretical computations, the hadronic contributions have been extracted (by assuming the SM)
from experimental data via dispersive integrals (HVP \& HLbL) and low-energy effective models supplemented with the
operator product expansion (HLbL). This leads to $a_\mu = 116 591 81.0(4.3)\times 10^{-10}$
(0.37 ppm)~\cite{Aoyama:2020ynm}, which deviates by $3-4$ standard deviations from the E821 result. That
difference has been persisting for a decade and it may be a hint for New Physics.

State-of-the-art lattice Quantum Chromodynamics (QCD) determinations of the HVP are becoming competitive.
At present, quoted uncertainties range between
$0.6\%$ to roughly $2\%$, see Ref.~\cite{Aoyama:2020ynm} and references therein, corresponding to an overall
error on $a_\mu$ which is still 5-15 times larger than the anticipated uncertainty
from E989. The main bottleneck~\cite{Aoyama:2020ynm} for matching that precision 
is the large statistical error in the Monte Carlo evaluation of the required correlation functions.
We have recently proposed~\cite{DallaBrida:2020cik} to solve this problem by a novel computational
paradigm based on multi-level Monte Carlo
integration in the presence of fermions~\cite{Ce:2016idq,Ce:2016ajy}. With respect to the
standard approach, this strategy reduces the variance exponentially with the temporal distance of the fields.
In this first feasibility study we focused on the HVP, but the strategy is general and it can be applied to the HLbL,
the isospin-breaking and electromagnetic contributions as well.
\vspace{-0.25cm}

\section{The signal-to-noise problem}
\vspace{-0.25cm}

The HVP can be written as
\be\label{eq:amuint}
a_\mu^{\rm HVP} = \left(\frac{\alpha}{\pi}\right)^2 \int_0^{\infty} d x_0\, K(x_0,m_\mu) \, G(x_0)\; ,
\ee
where $\alpha$ is the electromagnetic coupling constant, $K(x_0,m_\mu)$ is a known function
increasing quadratically at large $x_0$, $m_\mu$ is the muon mass,
and $G(x_0)$ is the zero-momentum correlation function 
\be\label{eq:correlat}
 G(x_0) = \int d^3 {\bf x}\, \langle J_k^{em}(x) J_k^{em}(0) \rangle
\ee
of two electromagnetic currents $J_k^{em}= i \sum_{i=1}^{N_f} q_i \bar\psi_i\gamma_k\psi_i$, for
unexplained notation see Ref.~\cite{DallaBrida:2020cik}. Here we consider $N_f=3$, the 3 lighter
quarks of QCD with the first 2 degenerate in mass, so that
\be
G(x_0) = G^{\rm conn}_{u,d}(x_0) + G^{\rm conn}_{s}(x_0) +  G^{\rm disc}_{u,d,s}(x_0)\; .  
\ee
The light-connected Wick contraction $G^{\rm conn}_{u,d}(x_0)$ and the disconnected
one $G^{\rm disc}_{u,d,s}(x_0)$ are the most problematic contributions with regard to the
statistical error. In standard Monte Carlo computations, the relative error of the former
at large time distances $|x_0|$ goes as  
\be\label{eq:relerr}
\frac{\sigma^2_{_{G^{\rm conn}_{\rm u,d}}}(x_0)}{[G^{{\rm conn}}_{\rm u,d}(x_0)]^2}\propto
\frac{1}{n_0}\; e^{2\, (M_\rho - M_\pi)|x_0|}\; ,  
\ee
where $M_\rho$ is the lightest asymptotic state in the iso-triplet vector
channel, and $n_0$ is the number of independent field configurations.
Therefore the computational effort, proportional to $n_0$, of reaching a given relative
statistical error increases exponentially with the distance $|x_0|$.
For the disconnected contribution $G^{\rm disc}_{u,d,s}(x_0)$, the situation is
worse since the variance is constant in time and therefore the coefficient multiplying
$|x_0|$ is larger. At present this exponential increase of the relative error is the barrier
which prevents lattice theorists to reach a per-mille statistical precision
for the HVP. 
\vspace{-0.25cm}

\section{Multi-level Monte Carlo}
\vspace{-0.25cm}

\begin{figure}[!t]
\vspace{-0.5cm}
  
\begin{center}
\includegraphics[width=9.0 cm,angle=0]{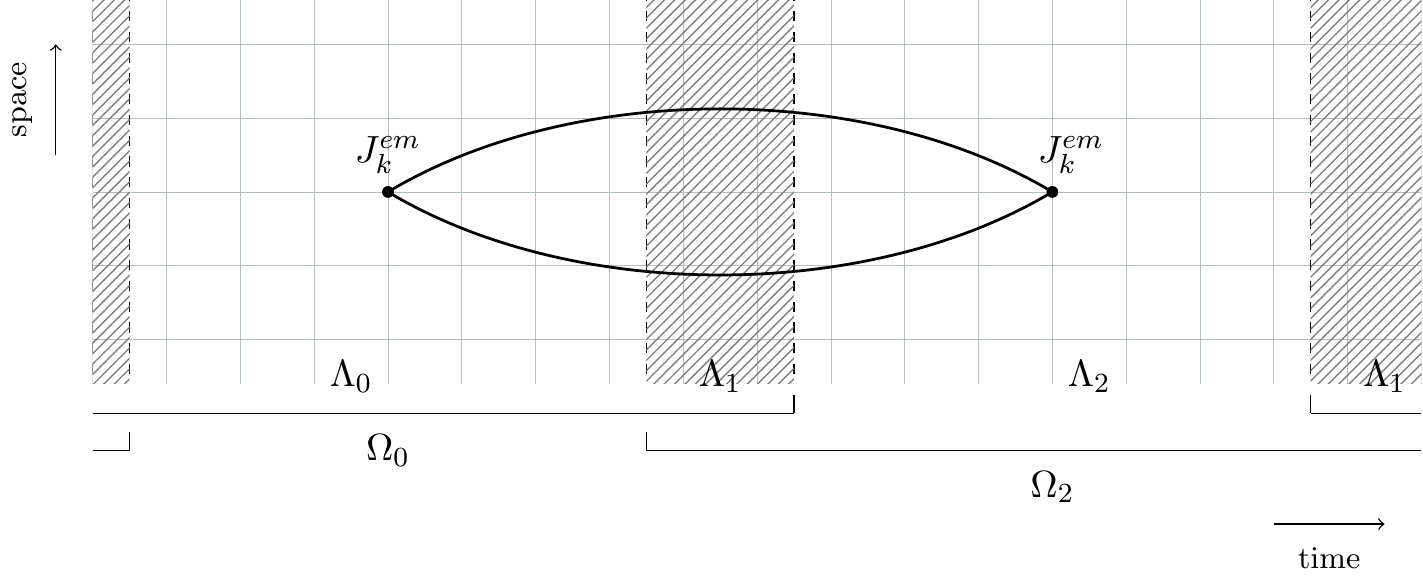}
\vspace{-0.25cm}

\caption{\label{fig:MB-DD} Domain decomposition of the lattice adopted here.
Periodic (anti-periodic) boundary conditions in the time direction are enforced for
gluons (fermions).}
\end{center}
\vspace{-1.0cm}

\end{figure}

Thanks to the conceptual, algorithmic and technical progress over the last few years,
it is now possible to carry out multi-level Monte Carlo simulations in the presence of
fermions~\cite{Ce:2016idq,Ce:2016ajy}. The first step in this approach is the decomposition of the
lattice in two overlapping domains $\Omega_0$ and $\Omega_2$, see e.g. Fig.~\ref{fig:MB-DD},
which share a common region $\Lambda_1$. The latter is chosen so that the minimum distance between the
points belonging to the inner domains $\Lambda_0$ and $\Lambda_2$ remains finite
and positive in the continuum limit. The next step consists in rewriting the
determinant of the Hermitean massive Wilson-Dirac
operator $Q=\gamma_5 D$ as
\begin{equation}\label{eq:factfinal0}
  \det\, Q= \frac{\det\, \left(1-w\right)}{\det\, Q_{\Lambda_{1}}
    \det\, Q^{-1}_{\Omega_0} \det\, Q^{-1}_{\Omega_2}} \; ,  
\end{equation}
where $Q_{\Lambda_{1}}$, $Q_{\Omega_0}$, and $Q_{\Omega_2}$ indicate the very same 
operator restricted to the domains specified by the subscript. They are obtained from $Q$
by imposing Dirichlet boundary conditions on the external boundaries of each domain. The 
matrix $w$ is built out of $Q_{\Omega_0}$, $Q_{\Omega_2}$ and the
hopping terms of the operator $Q$ across the boundaries in between the inner domains
$\Lambda_0$ and $\Lambda_2$ and the common region $\Lambda_1$~\cite{Ce:2016ajy}.
The denominator in Eq.~(\ref{eq:factfinal0}) has already a factorized
dependence on the gauge field since $\det Q_{\Lambda_{1}}$, $\det\, Q^{-1}_{\Omega_0}$
and $\det\, Q^{-1}_{\Omega_2}$ depend only on the gauge field in $\Lambda_1$, 
$\Omega_0$ and $\Omega_2$ respectively.
In the last step, the numerator in Eq.~(\ref{eq:factfinal0}) is rewritten as
\be\label{eq:MB}
\det\, \left(1-w\right) =
\frac{\det\,[1-R_{N+1}(1-w)]}{C \prod_{k=1}^{N/2} {\det} \big[ (u_k -w )^\dagger (u_k -w) \big]}\, ,
\ee
where $u_k$ and $u^*_k$ are the $N$ roots of a polynomial approximant for $(1-w)^{-1}$,
the numerator is the remainder, and $C$ is an irrelevant constant. The denominator
in Eq.~(\ref{eq:MB}) can be represented
by an integral over a set of $N/2$ multi-boson fields~\cite{DallaBrida:2020cik,Ce:2016idq,Ce:2016ajy}
having an action with a factorized dependence on the gauge field in $\Lambda_0$ and
$\Lambda_2$ inherited from $w$. When the polynomial approximation is properly chosen, the
remainder in the numerator of Eq.~(\ref{eq:MB}) has mild fluctuations in the gauge field, and it is
included in the observable in the form of a reweighting factor.

A simple implementation of these ideas is to divide the lattice as shown
in Fig.~\ref{fig:MB-DD}, where $\Lambda_0$ and $\Lambda_2$ have the shape of thick
time-slices while $\Lambda_1$ includes the remaining parts of the lattice.
The short-distance suppression of the quark propagator implies that
a thickness of $0.5$~fm or so for the thick-time slices forming $\Lambda_1$
is good enough, and is not expected to vary significantly with
the quark mass. This is the domain decomposition that we use for the numerical
computations  presented here.

The Monte Carlo simulation is then performed using a two-level scheme.
We first generate $n_0$ level-$0$ gauge field configurations by updating the field over the entire lattice;
then, starting from each level-$0$ configuration, we keep fixed the gauge field in the overlapping
region $\Lambda_1$, and generate $n_1$ level-$1$ configurations by updating 
the field in $\Lambda_0$ and in $\Lambda_2$ independently thanks to the factorization of the action.
The resulting gauge fields are then combined
to obtain effectively $n_0\cdot n_1^2$ configurations at the cost of generating $n_0\cdot n_1$
gauge fields over the entire lattice. Previous experience on two-level
integration suggests that, with two independently updated regions, the variance decreases proportionally
to $1/n_1^2$ until the standard deviation of the
estimator is comparable with the signal, i.e. until the level-$1$ integration has solved the signal-to-noise problem.
From Eq.~(\ref{eq:relerr}) we thus infer that the variance reduction due to level-$1$ integration
is expected to grow exponentially with the time-distance of the currents in Eq.~(\ref{eq:correlat}).
\vspace{-0.25cm}

\section{Lattice computation}
\vspace{-0.25cm}

In order to assess the efficiency of two-level Monte Carlo integration, we simulated QCD with two
dynamical flavours supplemented by a valence strange quark on a lattice of size $96\times 48^3$
with a spacing of $\;a=0.065$\,fm, and with a pion mass of $270$~MeV. The domains $\Lambda_0$ and
$\Lambda_2$ are the union of $40$ consecutive time-slices, while each thick time-slice forming
the overlapping region $\Lambda_1$ is made of $8$ time-slices. The determinants in the
denominator of Eq.~(\ref{eq:factfinal0}) are taken into account by standard pseudofermion
representations, while the number of multi-bosons is fixed to $N=12$. The very same action and
set of auxiliary fields are used either at level-$0$ or at level-$1$. The reweighting
factor is estimated stochastically with 2 random sources, which are enough for its contribution
to the statistical error to be negligible. We generate $n_0=25$ level-0 configurations, and
for each of them, we generate $n_1=10$ configurations in $\Lambda_0$ and $\Lambda_2$.
Further details on the algorithm and its implementation can be found in
Refs.~\cite{DallaBrida:2020cik,Ce:2016idq,Ce:2016ajy}.
\begin{figure*}[!t]
\vspace{-0.75cm}
  
\begin{center}
\begin{tabular}{cc}
\includegraphics[width=6.5 cm,angle=0]{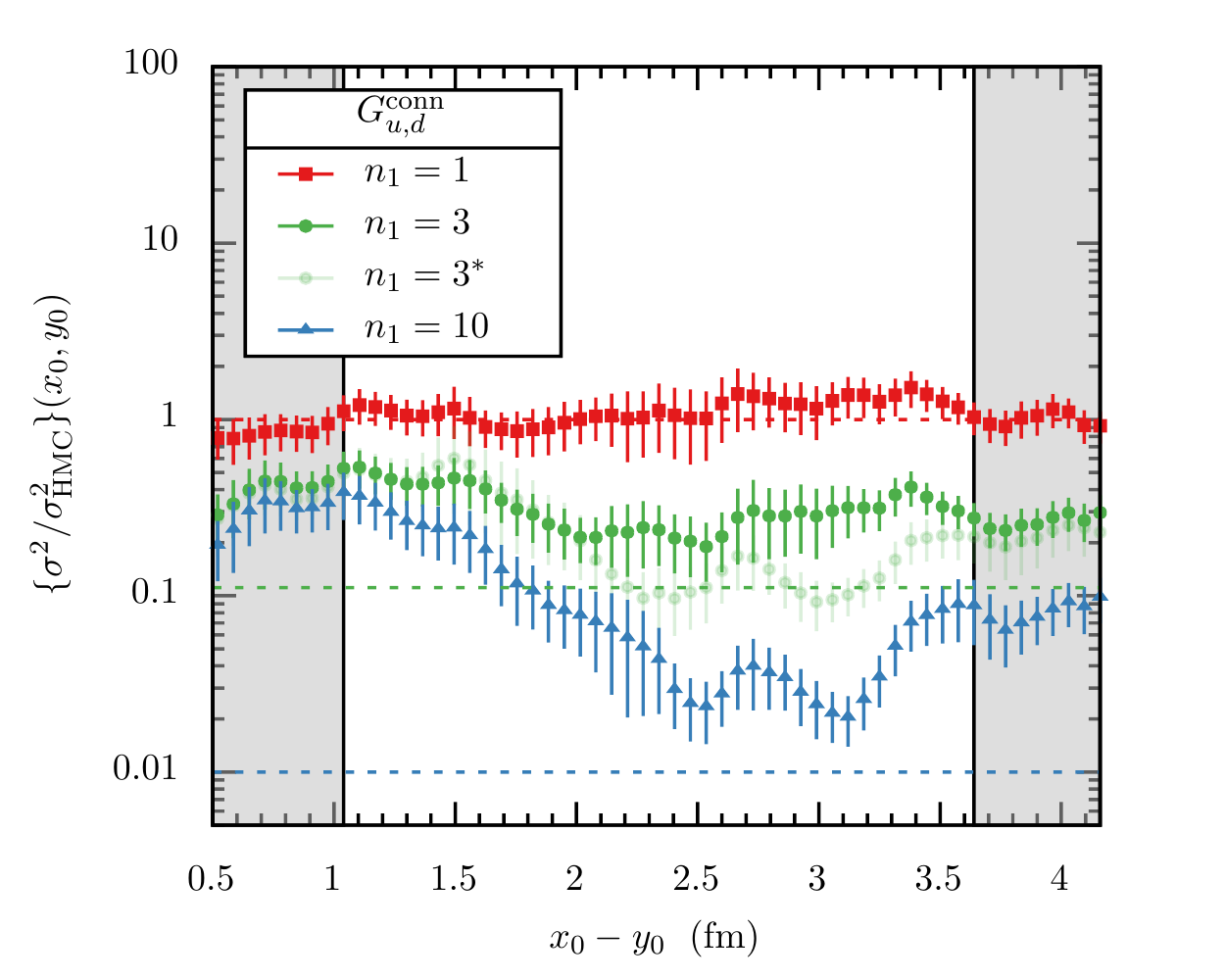} &
\includegraphics[width=6.5 cm,angle=0]{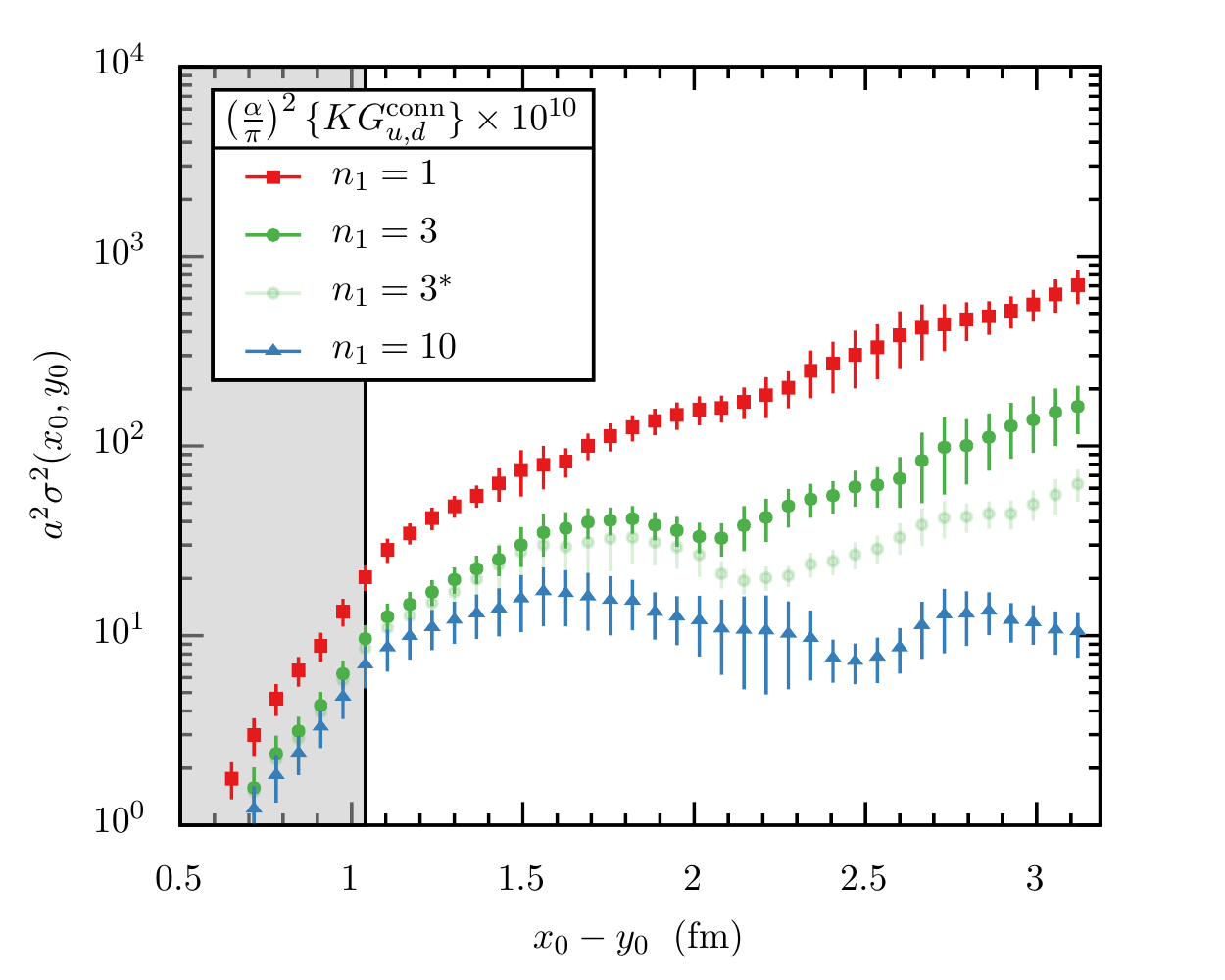}\\
\end{tabular}
\vspace{-0.5cm}

\caption{Left: variance of the light-connected contraction as a
function of the difference between the time-coordinates of the currents for
$n_1=1,3,10$. Data are normalized to the analogous ones computed on CLS
configurations generated by one-level HMC. Dashed lines represent the maximum reduction
which can be obtained by two-level integration, namely $1/n^2_1$, in the absence of
correlations between level-$1$ configurations. Grey bands indicate the thick time-slices
where the gauge field is kept fixed during level-$1$ updates. Right: variance of the light-connected
contribution to the integrand in Eq.~(\ref{eq:amuint}).
\label{fig:variances}}
\end{center}
\vspace{-0.875cm}

\end{figure*}
To single out the reduction of the variance due only to two-level averaging, we carry out a
dedicated calculation of correlation functions. We compute the light-connected contraction
by averaging over $216$ local sources put on the time-slice $y_0/a=32$ of
$\Lambda_0$ which is at a distance of $8$ lattice spacings from its right boundary 
and, as usual,
by summing over the sink space-position. We determine the disconnected contraction by
averaging each single-propagator trace over a large number of Gaussian random sources,
namely $768$, so to have a negligible random-noise contribution to the
variance~\cite{DallaBrida:2020cik,Giusti:2019kff}.

The variance of the light-connected contribution as a function of the distance from the source is shown
on the left plot of Fig.~\ref{fig:variances}. For better readability only the time-slices belonging
to $\Omega_2$ are shown, i.e. those relevant for studying the effect of two-level integration
given the source position. Data are normalized to the variance obtained with the same number of sources
on CLS configurations which were generated with a conventional one-level HMC.
The exponential reduction of the variance with the distance from the source is manifest in the data,
with the maximum gain reached from $2.5$~fm onward for $n_1=10$. The loss of about a factor between $2$ and
$3$ with respect to the best possible scaling, namely $n_1^2$, either for $n_1=3$ or $10$ (dashed lines) is
compatible with the presence of a residual correlation among level-$1$ configurations.
The power of the two-level integration can be better appreciated from the right plot
of Fig.~\ref{fig:variances},
where we show the variance of the light-connected contribution to the integrand in Eq.~(\ref{eq:amuint})
as a function of the time-distance of the currents. {The sharp rising of the variance}
computed by one-level Monte Carlo ($n_1=1$, red squares)
{is automatically flattened out by the two-level multi-boson domain-decomposed HMC} ($n_1=10$, blue triangles)
without the need for modeling the long-distance behaviour of $G^{\rm conn}_{u,d}(x_0)$.\\
\indent To further appreciate the effect of the two-level integration, we compute the integral in Eq.~(\ref{eq:amuint}) as
a function of the upper extrema of integration $x_0^{\rm max}$ which we allow to vary. For $n_1=1$, the
integral reads $446(26)$ and $424(38)$ for $x_0^{\rm max}=2.5$ and $3.0$~fm respectively, while
for $n_1=10$ the analogous values are $467.0(8.4)$ and $473.4(8.6)$. While with the one-level integration the errors
on the contributions to the integral from $0$ to $2.5$~fm and from $2.5$ to the maximum value of $3.0$~fm
are comparable, {with the two-level HMC the contribution to the variance from the long distance part becomes
negligible}. Considerations analogous to those made for the connected
contribution apply also to the much smaller disconnected one.
\vspace{-0.25cm}

\section{Results and discussion}
\vspace{-0.25cm}

\begin{figure*}[!t]
\vspace{-0.75cm}

\begin{center}
\begin{tabular}{cc}
\includegraphics[width=6.5 cm,angle=0]{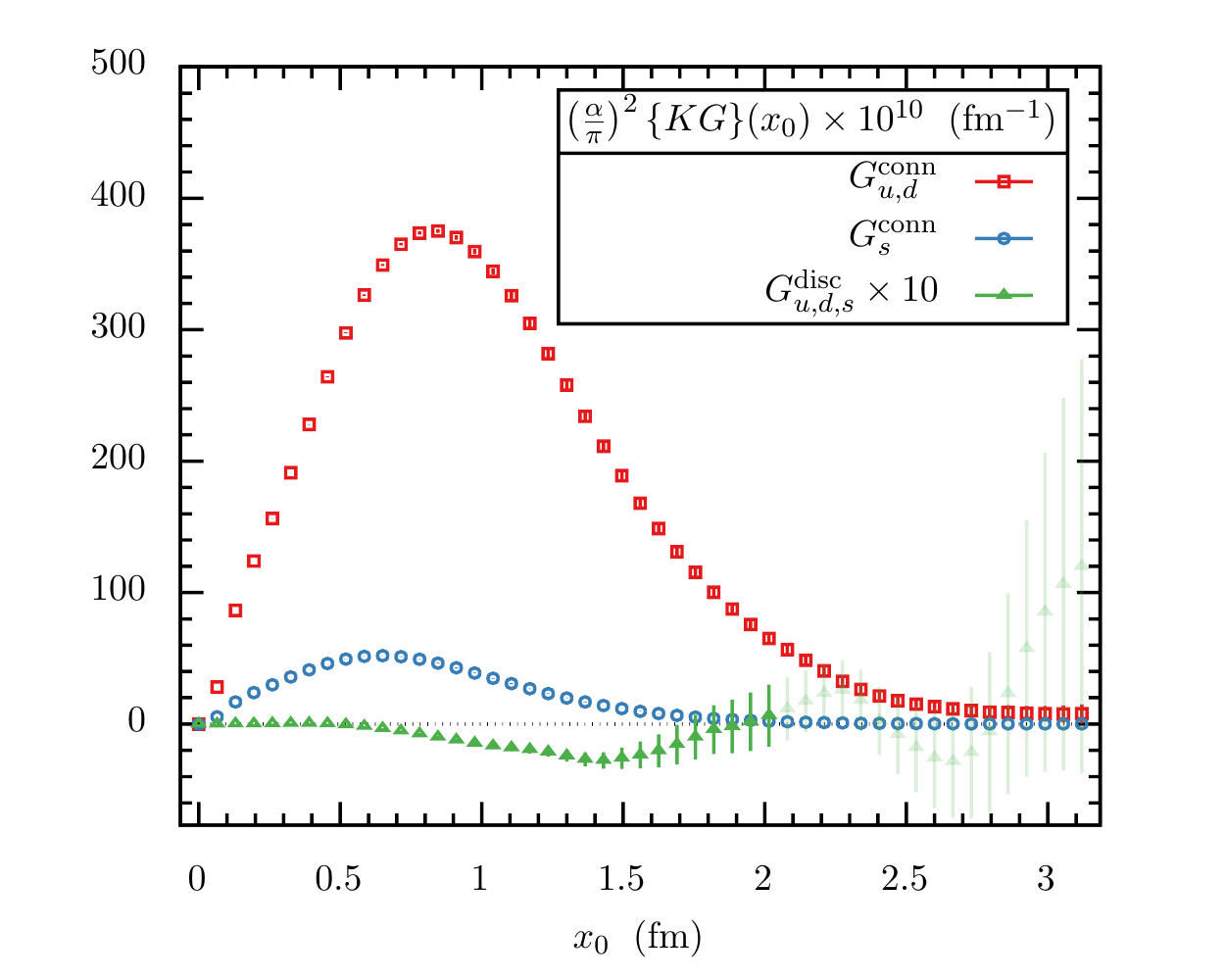} &
\includegraphics[width=6.5 cm,angle=0]{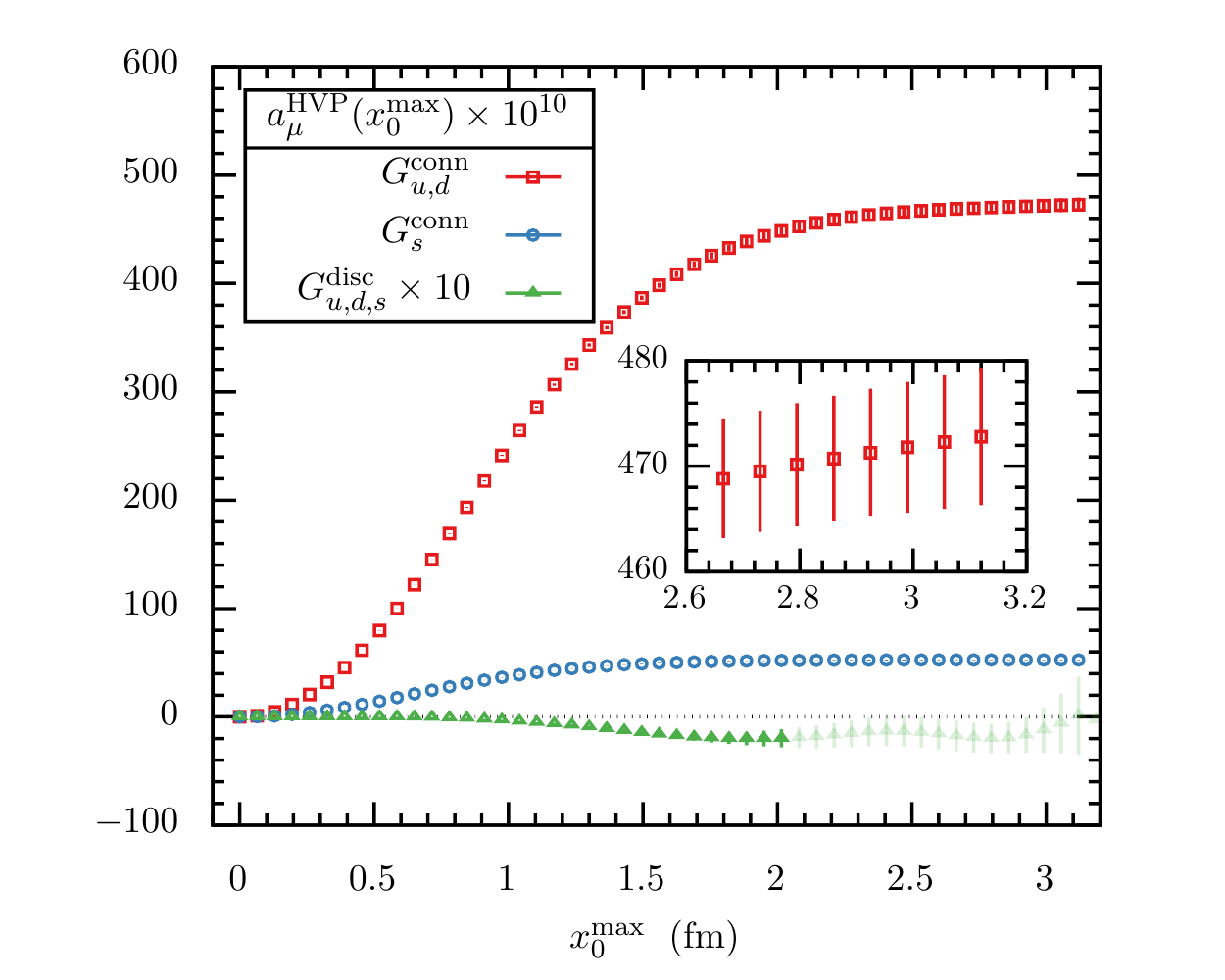}\\
\end{tabular}
\vspace{-0.5cm}

\caption{Left: best results for the contribution to the integrand in Eq.~(\ref{eq:amuint})
from the light-connected (red squares), strange-connected (blue circles) and disconnected (green triangles) contractions as
a function of the time coordinate. Right: best results for the contributions to
$a_\mu^{\rm HVP}$ from light-connected (red squares), strange-connected (blue circles), and disconnected (green triangles)
contractions as a function of $x_0^{\rm max}$.
\label{fig:integrand}}
\end{center}
\vspace{-0.875cm}

\end{figure*}

Our best result for the light-connected contribution to the integrand in Eq.~(\ref{eq:amuint})
is shown on the left plot of Fig.~\ref{fig:integrand} (red squares).
It is obtained
by a weighted average of the above discussed correlation function computed 
on $32$ point sources per time-slice on $7$ time-slices at $y_0/a=\{8,16,24,56,64,72,80\}$ and on $216$ sources at $y_0/a=32$.
We obtain a good statistical signal up to the maximum distance of $3$~fm or so.
The strange-connected contraction $G^{\rm conn}_{s}(x_0)$ is much less noisy, and it is determined
by averaging on $16$ point sources at $y_0/a=32$. Its value, shown on the left plot
of Fig.~\ref{fig:integrand} (blue circles), is at most one order of magnitude smaller than the
light contribution, and it has a negligible statistical error with respect to the light one.
The best result for the disconnected contribution has been computed as discussed in the previous section, and
it is shown in the left plot of Fig.~\ref{fig:integrand} as well (green triangles). It reaches a negative peak at
about $1.5$~fm, and a good statistical signal
is obtained up to $2.0$~fm or so. Its absolute value is more than two orders of magnitude smaller than the light-connected
contribution over the entire range explored.\\
\indent In the right plot of Fig.~\ref{fig:integrand} we show the best values of the light-connected (red squares),
strange-connected (blue circles), and disconnected (green triangles) contributions to $a_\mu^{\rm HVP}\cdot 10^{10}$
as a function of the upper extrema of integration $x_0^{\rm max}$ in Eq.~(\ref{eq:amuint}).
The light-connected part starts to flatten out at $x_0^{\rm max} \sim 2.5$~fm and, at the conservative distance of
$x_0^{\rm max}= 3.0$~fm, its value is $471.8(6.2)$. The value of the strange-connected contribution is $52.55(21)$
at $x_0^{\rm max}=3.0$~fm, and its error is negligible with respect to the light-connected one.
The disconnected contribution starts to flatten out at about $x_0^{\rm max} \sim 2.0$~fm, where its value
is $-1.98(84)$. For $x_0^{\rm max} =3.0$~fm, its statistical uncertainty is $2.1$ which is still
3 times smaller with respect to the light-connected one. Clearly the disconnected contribution must be taken
into account to attain a per-mille precision on the HVP, but the combined usage of split-even
estimators \cite{Giusti:2019kff} and two-level integration solves the problem of its computation. By combining the connected
contributions at $x_0^{\rm max}=3.0$~fm with the disconnected part at $x_0^{\rm max}= 2.0$~fm, the best total
value that we obtain is $a_\mu^{\rm HVP} = 522.4(6.2) \cdot 10^{-10}$.\\
\indent In this proof of concept study we have achieved a 1\% statistical precision with just
$n_0\cdot n_1=250$ configurations on a realistic lattice. This shows that for this light-quark mass
a per-mille statistical precision
on $a_\mu^{\rm HVP}$ is reachable with multi-level integration by increasing $n_0$ and $n_1$ by a factor of about
$4$--$6$ and $2$--$4$ respectively. When the up and the down quarks becomes lighter, the gain due
to the multi-level integration is expected to increase exponentially in the quark mass,
hence improving even more dramatically the scaling of the simulation cost with respect to a
standard one-level Monte Carlo. The change of computational paradigm presented here thus removes
the main barrier for making affordable, on computers 
available today, the goal of a per-mille precision on $a_\mu^{\rm HVP}$.

\end{document}